\documentclass[pra,twocolumn,showpacs,nobibnotes,floatfix,superscriptaddress]{revtex4}

\usepackage{latexsym,amsmath,amssymb,amsfonts,mathbbol,graphicx,color}
\usepackage{dcolumn}
\usepackage{bm}
\usepackage{graphicx}
\usepackage{epstopdf}

\begin{document}

\title{Non-Fault Tolerant $T$-Gates for the [7,1,3] Quantum Error Correction Code} 
\author{Yaakov S. Weinstein}
\affiliation{Quantum Information Science Group, {\sc Mitre},
200 Forrestal Rd. Princeton, NJ 08540, USA}

\begin{abstract}
We simulate the implementation of a $T$-gate, or $\frac{\pi}{8}$-gate, for a [7,1,3] encoded logical qubit in a non-equiprobable error environment. We demonstrate that the use of certain non-fault tolerant methods in the implementation may nevertheless enable reliable quantum computation while reducing basic resource consumption. Reliability is determined by calculating gate fidelities for the one-qubit logical gate. Specifically, we show that despite using a non-fault tolerant procedures in constructing a logical zero ancilla to implement the $T$-gate the gate fidelity of the logical gate, after perfect error correction, has no first order error terms. Meaning, any errors that may have occurred during implementation are `correctable' and fault tolerance may still be achieved.       
\end{abstract}

\pacs{03.67.Pp, 03.67.-a, 03.67.Lx}

\maketitle

\section{Introduction}
Quantum fault tolerance is a comprehensive framework which promises successful quantum computation despite errors to individual computational elements provided the error rate is below a certain threshold. This framework has been extensively researched over the past 15 years \cite{G, ShorQFT, Preskill, AGP, book} resulting in detailed rules on how to implement all elements of a quantum computation in a fault tolerant manner. One of the basic elements of quantum fault tolerance is to ensure that an error that occurs on one qubit cannot spread to multiple qubits. Application of quantum error correction (QEC) then corrects the single error \cite{book,ShorQEC,CSS}. Utilizing the entire framework of quantum fault tolerance in a practical quantum computation, however, promises to be a difficult and expensive proposition in terms of the number of physical qubits required and the number of physical gates implemented. Thus, it is worthwhile to explore the possibility of relaxing some of the strict rules required by the framework. 

The Calderbank-Shor-Steane (CSS) codes, a subclass of stabilizer quantum error correction codes, have proven to be very useful for the purposes of quantum fault tolerance. The reason for this is that Clifford gates can be performed in a bit-wise fashion. However, Clifford gates alone cannot be used to implement universal quantum computation. An additional gate such as the $\pi/8$-gate, also known as the $T$-gate, or Toffoli gate is necessary. These additional gates cannot be performed in a bit-wise fashion and thus turn out to the be the most difficult part of a fault tolerant quantum computation. In this paper we explore the implementation of the $T$-gate on a [7,1,3] quantum error correction code (or Steane code) \cite{Steane}, the most simple of the CSS codes. 

The $T$-gate is a single qubit $\pi/4$ phase shift with the matrix representation:
\begin{eqnarray}
T &=& 
\left( 
\begin{array}{cc}
1 & 0 \\
0 & e^{i\frac{\pi}{4}} \\
\end{array}
\right).
\end{eqnarray}
Its fault tolerant implementation for the [7,1,3] quantum error correction code requires an ancilla logical qubit in the state
\begin{equation}
|\Theta\rangle = \frac{1}{\sqrt{2}}(|0_L\rangle+e^{i\frac{\pi}{4}}|1_L\rangle,
\end{equation}
where $|0_L\rangle$ and $|1_L\rangle$ are the logical zero and one basis states. A controlled-NOT (CNOT) gate is then implemented between the ancilla and data qubits, the physical qubits storing the encoded logical qubit of information, with the ancilla as the control. The data qubits are measured and, if the measurement outcome is zero, the ancilla state is projected into the intial state of the data qubits with an applied $T$-gate. If the measurement outcome is a one, a NOT gate must be applied to the ancilla qubits to attain the desired outcome. The CNOT is a Clifford gate and can thus be applied bit-wise between the data and ancilla qubits. The NOT gate, if necessary, is also a Clifford gate. Thus, the most difficult part of implementing a the $T$-gate is the encoding of the state $|\Theta\rangle$. 

In this paper we analyze three methods of constructing the encoded state $|\Theta\rangle$ to see which can be used to implement usable $T$-gates for fault tolerant quantum computation. By `usable' we mean that the fidelity of the gate after application of perfect error correction has no first order error terms, i.e., all errors that occur during the gate are in principle correctable. The first method follows the tenets of fault tolerance as detailed in \cite{Preskill}. A logical state $|0_L\rangle$ is constructed via error correction on a state of all zeros. Measurement of the logical zero projects the qubits into $|\Theta\rangle$. Both the error correction and measurement are performed following the rules of fault tolerance. In addition, as part of the adherence to the rules of fault tolerance, the measurement is repeated until the same result is obtained twice in a row. The second method is to instead construct $|0_L\rangle$ via the encoding gate sequence of \cite{Steane}. This construction does not follow the rules of fault tolerance. Nevertheless, we have previously shown that gate encoded logical $|0_L\rangle$ states may be useable for quantum computation \cite{YSW} (see also \cite{BHW}). After the gate encoding, the logical zero state is projected into the desired state $|\Theta\rangle$ following the tenets of fault tolerance as per the first method. The third method is to use the gate encoding sequence to directly encode the single qubit state $|\theta\rangle = \frac{1}{\sqrt{2}}(|0\rangle+e^{i\frac{\pi}{4}}|1\rangle)$ into the state $|\Theta\rangle$. The first method describes a procedure which completely adheres to the rules of fault tolerance and thus the implemeted $T$-gate is expected to be usable for fault tolerant quantum computation. The second method does not follow fault tolerance procedure in constructing the logical zero but does follow them for measurement. The final method does not follow the rules of fault tolerance for any of its sub-protocols. We show that despite the lack of complete adherence to the rules of fault tolerance, the second method implements a logical $T$-gate with fidelities comparable to that of the first method (the `fault tolerant method') while the third method does not. Specifically, the gate fidelity of the logical $T$-gate implemented via the second method after perfect error correction does not have any first order error terms and should thus be usable for fault tolerant quantum computation. This implies that, in general, it may not be necessary to strictly and completely adhere to the tenets of quantum fault tolerance in order to implement fault tolerant quantum computation.

The error model used in this paper is a non-equiprobable Pauli operator error model \cite{QCC} with non-correlated errors. As in \cite{AP}, this model is a stochastic version of a biased noise model that can be formulated in terms of Hamiltonians coupling the system to an environment. In the model used here, however, the probabilities with which the different error types take place is left arbitrary: the environment causes qubits to undergo a $\sigma_x^j$ error with probability $p_x$, a $\sigma_y^j$ error with probability $p_y$, and a $\sigma_z^j$ error with probability $p_z$, where $\sigma_i^j$, $i = x,y,z$ are the Pauli spin operators on qubit $j$. For example, a single qubit gate, $U_j$ performed in such an environment on a qubit $j$ in the state $\rho_j$ undergoes the following evolution: 
\begin{equation}
\sum_{a=0,x,y,z}p_a\sigma_a^jU_j\rho_j U_j^{\dag}\sigma_a^j,
\end{equation} 
where $\sigma_0^j$ is the identity matrix, $p_0 = 1-\sum_{\ell=x,y,z}p_\ell$, and the terms $K_a^j = \sqrt{p_a}\sigma_a^jU_j$ can be regarded as Kraus operators for the 
single qubit gate. Similarly, a two-qubit gate, $U_{j,k}$ implemented in this environment on a two qubit state $\rho_{j,k}$ actually implements: 
\begin{equation}
\sum_{a,b=0,x,y,z}p_ap_b\sigma_a^j\sigma_b^k U_{j,k} \rho_{j,k} U_{j,k}^{\dag}\sigma_a^j\sigma_b^k,
\label{cnot}
\end{equation}
where terms $A_{a,b}^{j,k} = \sqrt{p_ap_b}\sigma_a^j\sigma_b^k U_{j,k}$ can be regarded as the 16 Kraus operators. Note that errors on the two qubits taking part in the two qubit gate are independent and not correlated. We assume that only qubits taking part in a gate operation will be subject to error. Qubits not involved in a gate are assumed to be perfectly stored. While this represents an idealization, it is partially justified in that
it is generally assumed that stored qubits are less likely to undergo error than those involved in gates (see for example \cite{Svore}). In addition, in this paper accuracy measures are calculated only to second order in the error probabilities $p_i$ thus the effect of ignoring storage errors is likely minimal. Finally, we note that non-equiprobable errors occur in the initialization of qubits to the $|0\rangle$ state and measurement (in the $z$ or $x$ bases) of all qubits.

\section{$T$-Gate Implementations}

The method for encoding a logical zero state in the Steane code following the rules of fault tolerance is to apply error correction following the rules of fault tolerance to 7 qubits all initially in the state $|0\rangle$ \cite{Preskill}. Were the initialization perfect there would be no need to perform the bit-flip syndrome measurements as they will have no effect on the state of the qubits. Due to the non-equiprobable error environment, however, the initial state of the qubits will not be $|0\rangle$ but $\rho_i=(1-p_x-p_y)|0\rangle\langle 0|+(p_x+p_y)|1\rangle\langle 1|$. Neverthless, due to risk of doing more harm than good, we choose not to apply the bit-flip syndrome measurements, and instead apply the phase-flip syndrome measurements only (this is done twice to conform with the strictures of fault tolerance). We analyze the scenario where all syndrome measurements yield a zero. Because encoding is done `off-line' one can choose to utilize only the encoded states with this outcome. 

\begin{figure}
\includegraphics[width=8.5cm]{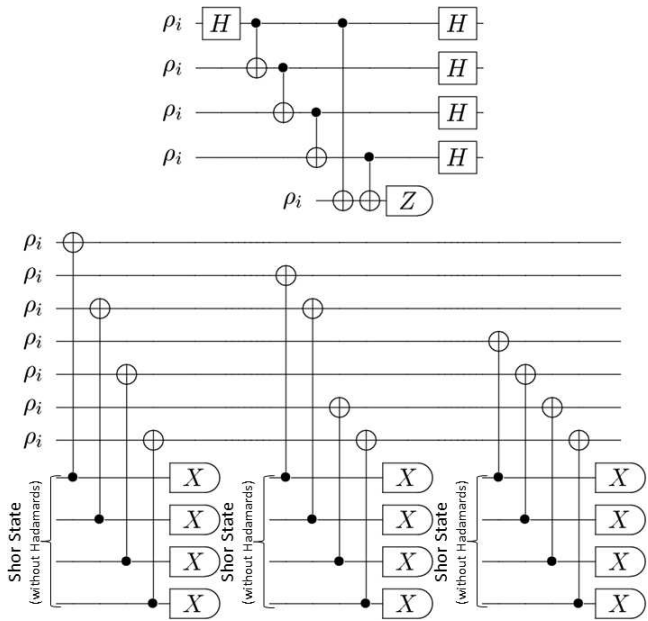}
\caption{Top: construction of a 4 qubit Shor state. \textsc{cnot} gates are represented by ($\bullet$) on the control qubit and ($\oplus$) on the target qubit connected by a vertical line. $H$ represents a Hadamard gate. The procedure entails constructing a GHZ state which is verified using an ancilla qubit. Hadamard gates are applied to each qubit to complete Shor state construction.
Bottom: phase-flip syndrome measurements for the [7,1,3] code following fault tolerance procedures using Shor states. To ahere to the tenets of fault tolerance, each Shor state ancilla qubit must interact with only one data qubit. The error syndrome is determined from the parity of the measurement outcomes of the Shor state ancilla qubits. Note that the Shor states utilized are without the final Hadamard gates and thus we reverse the roles of the control and target qubits and measure the ancilla in the $x$-basis as explained in \cite{Preskill}. Following fault tolerance procedure each of the syndrome measurements is repeated twice.}
\label{FT}
\end{figure}

Error correction following the rules of fault tolerance requires proper ancilla qubits to determine the syndrome measurement. We choose four-qubit Shor states \cite{ShorQFT} for ancilla as they require the least number of qubits and are thus most likely to be experimentally accessible in the short term. Shor states are simply Greenberger-Horne-Zeilinger (GHZ) states with Hadamard gates applied to each qubit. Because the Shor states themselves are constructed in a noisy environment (here the nonequiprobable error environment), verification via parity checks on pairs of qubits is necessary to ensure accurate construction. In consonance with the results of \cite{WB} we apply one verification. Circuits for Shor state construction and verification, and the circuit for the three bit-flip syndrome measurements used to encode the logical zero state are shown in Fig.~\ref{FT}. 

\begin{figure}
\includegraphics[width=8cm]{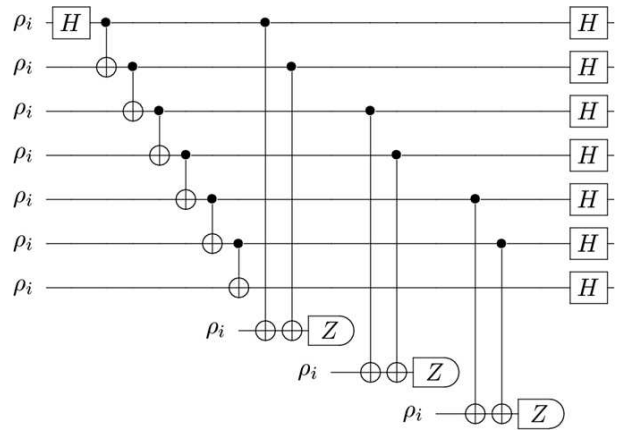}
\caption{Construction of a 7-qubit Shor state with three verifications. This Shor state can then be used to project the logical zero state into the logical state $|\Theta\rangle$. }
\label{Shor7}
\end{figure}

After constructing a logical zero state we are ready to project into the state $|\Theta\rangle$. To do this following the rules of fault tolerance we need a seven qubit Shor state. The Shor state construction, shown in Fig.~\ref{Shor7}, is done in the non-equiprobable error environment and employs three verification steps. We then apply controlled-M gates given by: 
\begin{eqnarray}
CM &=& 
\left( 
\begin{array}{cccc}
1 & 0 & 0 & 0 \\
0 & 1 & 0 & 0 \\
0 & 0 & 0 & e^{i\frac{\pi}{4}} \\
0 & 0 & e^{-i\frac{\pi}{4}} & 0 \\
\end{array}
\right)
\end{eqnarray}
with the Shor state qubits as control and the logical zero state qubits as targets. Measurement of the Shor state (with even parity outcome) completes the projection and the construction of the logical state $|\Theta\rangle$. The entire procedure is done until the same syndrome is attained twice in a row to ensure no errors have taken places during the projection.  

\begin{figure}
\includegraphics[width=5.5cm]{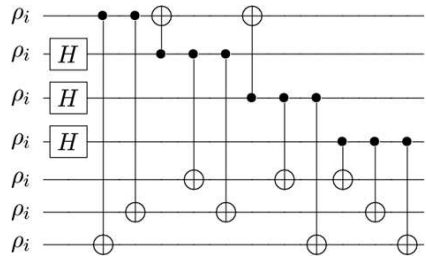}
\caption{Encoding of logical zero state via the gate encoding sequence. This method of encoding does not follow the rules of fault tolerance.}
\label{GSEnc}
\end{figure}

The second method we use to construct the logical $|\Theta\rangle$ is the same as the fault tolerant method except that instead of using a logical zero state encoded via the rules of fault tolerance we use a logical zero constructed from the encoding gate sequence of Ref.~\cite{Steane} shown in Fig.~\ref{GSEnc}. This construction does not follow the tenets of fault tolerance; an error on one qubit can easily spread to other qubits. 

The final method used to construct the logical $|\Theta\rangle$ is to directly encode the one qubit state $|\theta\rangle$, which requires applying a Hadamard gate and a $T$-gate to the first qubit and then follow the gate encoding sequence above. Again this method does not follow the rules of fault tolerance. However, it is the shortest and most direct method of constructing the logical $|\Theta\rangle$ state. We have simulated all of these construction methods in the non-equiprobable error environment described above. 

After construction of the logical $|\Theta\rangle$ we apply a CNOT gate between it and a perfectly encoded arbitrary state $|\psi_L\rangle=\cos\alpha|0_L\rangle+e^{i\beta}\sin\alpha|1_L\rangle$. The CNOT is implemented in the non-equiprobable Pauli error environment. The encoded arbitrary state is then (noisily) measured and, assuming the measurement outcomes have even parity the output of the $T$-gate on the arbitrary encoded state is found on the qubits initially in the logical $|\Theta\rangle$ state (if the measurement outcomes yield odd parity a NOT gate must be applied). We will eventually replace the arbitrary state with the four states necessary to simulate one qubit quantum process tomography. This will allow us to calculate a logical gate $\chi$-matrix.

\section{Results}

We invoke a number of accuracy measures to compare the three $T$-gate implementation methods. The goal of these measures is to determine the possibility of using the various $T$-gate implementations in fault tolerant quantum computation. The first accuracy measures are the fidelity of the constructed seven-qubit $|\Theta\rangle$ state $\langle\Theta|\rho_{\Theta-sim}|\Theta\rangle$, where $\rho_{\Theta-sim}$ is the state resulting from simulating our $|\Theta\rangle$ state construction methods, and the fidelity of the one-qubit logical state $|\theta\rangle$ calculated by perfectly decoding $\rho_{\Theta-sim}$, given by $\langle\theta|\rho_{\theta-sim}|\theta\rangle$. These fidelities, shown in Table \ref{Tab1}, suggest how well the $|\Theta\rangle$ state construction process is carried out but may not indicate how accurately the $T$-gate utilizing this state will perform. A couple of interesting points are demonstrated by the fidelity results. First, the third method, in which no part of the state construction follows the rules of fault tolerance, in general produces higher fidelity $|\Theta\rangle$ states than the other two methods. Second, when implementing either of the first two methods the $\sigma_z$ errors are significantly less important than the other two error types. This is not true for the third method. 

\begingroup
\squeezetable
\begin{table*}
\caption{Fidelity measures to first order in error probabilities  for the seven and one-qubit state $|\theta\rangle$ for the three $|\theta\rangle$-construction methods: fault tolerant construction, construction using a gate encoded logical zero state, and direct gate encoding construction. }
\begin{tabular}{||c||c|c|c||}
\hline 
 & fault tolerant & gate encoded $|0\rangle$ & gate encoded $|\Theta\rangle$ \\\hline
\hline
7-Qubit fidelity  & $1-\frac{167p_x}{2}-\frac{71p_y}{2}-19p_z$  & $1-\frac{125p_x}{2}-\frac{95p_y}{2}-28p_z$ & $1-\frac{59p_x}{2}-\frac{65p_y}{2}-26p_z$   \\\hline
1-Qubit fidelity  & $1-\frac{149p_x}{4}-\frac{69p_y}{4}-11p_z$ & $1-\frac{113p_x}{4}-\frac{105p_y}{4}-20p_z$ & $1-\frac{9p_x}{2}-\frac{21p_y}{2}-11p_z$  \\\hline
\end{tabular}
\label{Tab1}
\end{table*}
\endgroup

Our goal in this work is not simply to construct $|\Theta\rangle$ states but to use them to implement a $T$-gate. The fact that the $|\Theta\rangle$ state constructed from the third method has the highest fidelity does not necessarily mean it will provide for the most accurate $T$-gates. To check this we now turn to accuracy measures after $T$-gate implementation. These are: the seven-qubit and one logical qubit output state of a $T$-gate applied to a perfectly encoded arbitrary state $|\psi_L\rangle$, and the gate fidelity of the one-qubit logical $T$-gate. The gate fidelity is determined by constructing a $\chi$-matrix for the one-qubit logical gate. Each of these fidelity measures is determined at three different computational points: after implementation of the $T$-gate, after implementation of the $T$-gate and perfect error correction, and after implementation of the $T$-gate and error correction done in the non-equiprobable Pauli error environment. We stress that it is not sufficient to simply look at accuracy measures of the $T$-gate alone. Error correction is an integral part of fault tolerant quantum computation and may need to be applied after every gate. Applying perfect error correction allows us to test the `correctability' of the errors that occur during $T$-gate implementation. If even perfect error correction cannot (to first order) correct the errors in the $T$-gate then it is unlikely that this gate can be used for practical implementations of quantum computation. Noisy error correction, on the other hand, tells us practically how well a $T$-gate with error correction can be performed. The output state fidelities for an arbitary input state (for convenience we will call this the arbitrary output state) after these processes are given in Table \ref{Tab2}.  

\begingroup
\squeezetable
\begin{table*}
\caption{Fidelity measures to first order in error probabilities  for the seven and one-qubit output states after application of the $T$-gate, the $T$-gate and perfect error correction, and the $T$-gate and noisy error correction on an arbitrary input state.}
\begin{tabular}{||c||c|c|c||}
\hline 
7-qubit & fault tolerant & gate encoded $|0\rangle$ & gate encoded $|\theta\rangle$ \\\hline
\hline
$T$-gate  & $1-7p_x-7p_y-26p_z$ & $1-10p_x-13p_y-35p_z$ & $1-\frac{43p_x}{4}-\frac{55p_y}{4}-\frac{61p_z}{2}$ \\\hline
$T$ + perfect QEC  & $1$ & $1$ & $1-(\frac{3p_x-3p_y-10p_z}{4})(1-\cos[4a])$\\\hline
$T$ + noisy QEC  & $1-73p_x-19p_y-7p_z$ & $1-73p_x-19p_y-7p_z$ & $1-\frac{295p_x}{4}-\frac{79p_y}{4}-\frac{19p_z}{2}$ \\
& & & $+\frac{3(p_x+p_y)+10p_z}{4}\cos[4a]$ \\\hline \hline
logical qubit & fault tolerant & gate encoded $|0\rangle$ & gate encoded $|\theta\rangle$ \\\hline
\hline
$T$-gate  & $1-\frac{3p_x}{4}(3+\cos[4a])$ & $1-\frac{3p_x}{4}(5+\cos[4a])$ & $1-3p_x-p_y(4-\cos[4a])$ \\
 & $-\frac{p_y}{4}(13-\cos[4a])-7p_z(1-\cos[4a])$ & $-\frac{p_y}{4}(25-13\cos[4a])-\frac{23p_z}{2}(1-\cos[4a])$ & $-7p_z(1-\cos[4a])$ \\
& & & $+\frac{3p_x+p_y}{2}\sin[2a]^2\sin[2b]$\\\hline
$T$ + perfect QEC  & $1$ & $1$ & $1-\frac{3(p_x+p_y)}{4}(1-\cos[4a])$ \\
 & & & $-\frac{5p_z}{2}(1-\cos[4a])$ \\\hline
$T$ + noisy QEC  & $1-\frac{19p_x}{4}(3+\cos[4a])$ & $1-\frac{19p_x}{4}(3+\cos[4a])$ & $1-p_x(15-4\cos[4a])$ \\
& $-\frac{p_y}{4}(13-\cos[4a])-\frac{3p_z}{2}(1-\cos[4a])$ & $-\frac{p_y}{4}(13-\cos[4a])-\frac{3p_z}{2}(1-\cos[4a])$ & $-p_y(4-\cos[4a])-4p_z(1-\cos[4a])$\\ 
& & & $+\frac{19p_x+p_y}{2}\sin[2a]^2\sin[2b]$ \\\hline
\end{tabular}
\label{Tab2}
\end{table*}
\endgroup

The output state fidelities of the $T$-gate alone are comparable for all three methods with the $\sigma_z$ errors being dominant. However, after perfect error correction it becomes clear that only the fault tolerant method and the gate-encoded logical zero method give output states that are acceptable for practical quantum computation. This is because whatever errors may have arisen during the implementation are `correctable' in that they are suppressed to second order in the fidelity by the perfect error correction. This is not true of the gate-encoded $|\Theta\rangle$ state method and the $T$-gate implemented via this method must be regarded as not fault tolerant and unacceptable for quantum computation. 

If we were to apply noisy error correction after implementation of the $T$-gate we could gauge how accurately an actual complete process will be performed. Noisy error correction is applied using (noisy) four-qubit Shor states as ancilla for syndrome measurement. Each syndrome measurement is repeated until the same result is obtained twice in a row. We have simulated the case where all syndrome measurements give a zero, as explained in \cite{WB}. After noisy error correction the fidelity of the output states for the first two methods are exactly the same up to first order (and the second order terms are also nearly the same). This implies that the most recent operation is what drives the first order fidelity terms. The fact that an earlier part of the $T$-gate implementation did not adhere to the rules of fault tolerance seems to have been washed away. We also note that the $\sigma_x$ errors become dominant. As explained in \cite{WB}, this is primarily due to two reasons, the fact that the bit-flip syndrome measurements are done first, and the use of imperfect Shor ancilla states.

Once the general arbitrary output state of the $T$-gate is calculated it is straightforward to construct the one logical qubit $\chi$-matrix following the procedure outlined in \cite{QPT,book}. The gate fidelity of the $T$-gate is then calculated from the $\chi$-matrix as ${\rm{Tr}}[\chi(p_j)\chi(0)]$, where $\chi(p_j)$ is the simulated $\chi$-matrix of the $T$-gate implemented in the non-equiprobable error environment with error probabilities $p_j$. We calculate gate fidelities after implementation of the logical $T$-gate, the logical $T$-gate and perfect error correction, and the logical $T$-gate and noisy error correction. The results are shown in Table \ref{Tab3}.  

\begingroup
\squeezetable
\begin{table*}
\caption{Gate fidelity measures calculated from single logical qubit $\chi$-matrices to first order in error probability for the $T$-gate, the $T$-gate and perfect error correction, and the $T$-gate and noisy error correction.}
\begin{tabular}{||c||c|c|c||}
\hline 
Gate Fidelity & fault tolerant & gate encoded $|0\rangle$ & gate encoded $|\theta\rangle$ \\\hline
\hline
$T$-gate  & $1-3p_x-5p_y-14p_z$ & $1-6p_x-11p_y-23p_z$ & $1-14p_x-27p_y-52p_z$ \\\hline
$T$ + perfect QEC  & $1$ & $1$ & $1-\frac{3(p_x+p_y)}{2}-5p_z$ \\\hline
$T$ + noisy QEC  & $1-19p_x-5p_y-3p_z$ & $1-19p_x-5p_y-3p_z$ & $1-\frac{41p_x}{2}-\frac{13p_y}{2}-8p_z$ \\\hline 
\end{tabular}
\label{Tab3}
\end{table*}
\endgroup

The gate fidelities follow the same trends found in arbitrary output state fidelities. They demonstrate the usability for quantum computation of not only the fault tolerant method $T$-gate but also the $T$-gate implemented using a gate encoded logical zero state. Finally, we note that the fidelity after noisy error correction is not necessarily better than the fidelity before noisy error correction. This should not be taken to mean that error correction is not necessary. We have already seen that fidelity at one stage of a computation does not translate into optimum performance at a later stage of the computation. The most we can say is that this may suggest that error correction need not be applied after every computational step. This will be explored in future work.  

\section{Discussion and Conclusions}

The simulations presented in this work shed light on a number of issues related to fault tolerance. First, we question whether every element of a computation must indeed be constructed in a manner consistent with the rules of fault tolerance as outlined, for example, in \cite{Preskill} to be usable for practical quantum computation. We define a `useable' operation as one that, after perfect error correction, has no remaining first order error probability terms in fidelity. The $T$-gate is explored since, as a non-Clifford gate, it is the most difficult to implement for a CSS code and is therefore most in need of shortcuts. We have found that the logical zero ancilla called for need not be constructed following the tenets of fault tolerance and that gate sequence encoded logical zero will yield a usable $T$-gate. This provides a significant savings of number of gates, time of operation, and qubits. However, a direct gate encoding of the state $|\Theta\rangle$ will not yield a usable operation. Additional simulations demonstrate that another attempted shortcut, applying the projection of the logical zero state into the $|\Theta\rangle$ state only once and not repeating it to attain the same measurement results twice in a row, will also not yield a usable $T$-gate. 

Throughout we have looked at fidelities of the gate implementation and in this way determined if a gate is `usable.'  Are usable gates fault tolerant? Can they be used for arbitrarily long computations without undue build up of errors? Though we cannot prove equivalence between usable and fault tolerant there is evidence implying that this is the case. First, when perfect error correction is applied to a usable gate it achieves perfect fidelity (to at least second order). This implies that the gate is fault tolerant as error correction will fix all errors. Second, upon noisy error correction the usable gate has a fidelity equivalent to the gate implemented via the fault tolerant method implying that any errors from previous protocols will be washed out by later ones and again that the gate is fault tolerant. 

A second issue is the need to perform error correction after every step in a computation. While this issue must be fully addressed elsewhere we would like to make three points here. First, in the fault tolerant method we did not apply the bit-flip syndrome measurements for logical zero state encoding even though initialization was peformed imperfectly. This lack does not appear to have negatively affected any results. The fault tolerant method still yielded usable gates. Second, we did not apply quantum error correction to the logical zero states in either the fault tolerant method or gate-encoded logical zero method and still the logical zero states led to implementing useable $T$-gates. Third, applying realistic (noisy) error correction after implementation of the $T$-gate has not improved the fidelity of the operation. If anything it makes it worse. Perhaps, applying a few operations before error correction would not severely harm the fidelity of the operations. 

Thirdly, we would like to point out the utility of the logical $\chi$-matrix in evaluating the accuracy of the $T$-gate performance. The $\chi$-matrix is easily transformed into Kraus operators which properly describe the one qubit sequence: perfect encoding, implementation of $T$-gate, perfect decoding. Such Kraus operators may be useful for simulations of quantum fault tolerance. 

In conclusion, we have explored the possibility of utilizing non-fault tolerant methods to implement $T$-gates for fault tolerant quantum computation. We have shown that when certain elements of fault tolerant protocols are relaxed, the $T$-gate can still be implemented in such a way such that (ideal) error correction would correct errors to first order (in the fidelity). Relaxing other elements of fault tolerance, however, would cause the gate to be unusable. Further work is necessary to outline a general model as to what elements are `non-essential' in this way. While this work was done in the context of the [7,1,3] quantum error correction code we believe it would be immediately applicable to other CSS codes and possibly to more general codes. Our work also has implications for the question of how often error correction must be applied during a fault tolerant quantum computation, and we have begun to explore the utility of logical qubit Kraus operators for fault tolerant simulations.    

This research is supported under MITRE Innovation Program Grant 51MSR662.


\begin{thebibliography}{99}

\bibitem{Preskill}
J. Preskill, Proc. Roy. Soc. Lond. A {\bf 454}, 385 (1998).
\bibitem{ShorQFT}
P.W. Shor, {\it Proceedings of the the 35th Annual Symposium on Fundamentals of Computer Science}, (IEEE Press, Los Alamitos, CA, 1996).
\bibitem{G}
D. Gottesman, Phys. Rev. A {\bf 57}, 127 (1998).
\bibitem{AGP}
P. Aleferis, D. Gottesman, and J. Preskill, Quant. Inf. Comput. {\bf 6}, 97 (2006).
\bibitem{book}
M. Nielsen and I. Chuang, {\it Quantum information and Computation} (Cambridge University Press, Cambridge, 2000).
\bibitem{ShorQEC}
P.W. Shor, Phys. Rev. A {\bf 52}, R2493 (1995).
\bibitem{CSS}
A.R. Calderbank and P.W. Shor, Phys. Rev. A {\bf 54}, 1098 (1996);
A.M. Steane, Phys. Rev. Lett. {\bf 77}, 793 (1996).
\bibitem{Steane}
A.M. Steane, Proc. Roy. Soc. Lond. A {\bf 452}, 2551 (1996).
\bibitem{YSW}
Y.S. Weinstein, Phys. Rev. A {\bf 84}, 012323 (2011).
\bibitem{BHW}
S.D. Buchbinder, C.L. Huang, and Y.S. Weinstein, Quant. Inf. Proc. {\bf 12}, 699 (2013).
\bibitem{QCC}
V. Aggarwal, A.R. Calderbank, G. Gilbert, Y.S. Weinstein, Quant. Inf. Proc. {\bf 9}, 541 (2010).
\bibitem{AP}
P. Aliferis and J. Preskill, Phys. Rev. A {\bf 78}, 052331 (2008).
\bibitem{Svore}
K.M. Svore, B.M. Terhal, D.P. DiVincenzo, Phys. Rev. A {\bf 72}, 022317 (2005). 
\bibitem{WB}
Y.S. Weinstein and S.D. Buchbinder, Phys. Rev. A {\bf 86}, 052336 (2012).
\bibitem{QPT}
I.L. Chuang and M.A. Nielsen J. Mod. Opt. {\bf 44}, 2455 (1997). 

\end{thebibliography}
\end{document}